\documentclass[sigconf]{acmart}

\setcopyright{none}
\renewcommand\footnotetextcopyrightpermission[1]{}

\usepackage{fancyhdr}
\fancypagestyle{plain}{%
   \fancyhf{} %
   \fancyfoot[L]{ACM SIGCOMM Computer Communication Review }%
   \fancyfoot[R]{Volume 50 Issue 2, April 2020}%
}
\def\refname{}
\usepackage{balance}
\usepackage{xspace}
\usepackage{graphicx}
\usepackage{caption}
\usepackage{subcaption}
\usepackage[export]{adjustbox}

\newcommand{\xref}[1]{\S\ref{#1}}

\newcommand{\cref}[1]{Chapter~\ref{#1}}

\newcommand{\fref}[1]{Fig.~\ref{#1}}

\newcommand{\ie}{i.e., \@}
\newcommand{\eg}{e.g., \@}

\definecolor{darkgreen}{rgb}{0,0.5,0}
\definecolor{brown}{rgb}{0.7,0.3,0}
\definecolor{darkblue}{rgb}{0,0,0.5}
\definecolor{darkred}{rgb}{0.5,0,0}

\newcounter{fn1}
\newcounter{fn2}
\newcounter{fn3}
\newcounter{fn4}
\newcounter{fn5}
\newcommand{\myitem}[1]{\vspace*{0.03in}\noindent\emph{\textbf{#1}}\enskip}

\newcommand{\remove}[1]{}
\newcommand{\myitemit}[1]{\vspace*{0.02in}\noindent\emph{\textit{#1}}\enskip}

\newcommand{\name}{mini-Internet\xspace}

\newcommand{\shellcmd}[1]{\vspace*{0.07in}\noindent\texttt{\textbf{>} #1}\vspace*{0.01in}}
\newcommand{\shellcmdproxy}[1]{\vspace*{0.01in}\noindent\texttt{\textbf{g1-proxy>} #1}\vspace*{0.01in}}
\newcommand{\routercmd}[1]{\vspace*{0.01in}\noindent\texttt{\textbf{3-router\#} #1}\vspace*{0.07in}}

\begin{document}

\title{An Open Platform to Teach How the Internet Practically Works}

\subtitle{\href{http://mini-inter.net}{\LARGE\textcolor{blue}{\textbf{\texttt{mini-inter.net}}}}}

\author{Thomas Holterbach}
\affiliation{
	\institution{ETH Zurich}
}
\email{thomahol@ethz.ch}

\author{Tobias B{\"u}hler}
\affiliation{
	\institution{ETH Zurich}
}
\email{buehlert@ethz.ch}

\author{Tino Rellstab}
\affiliation{
	\institution{ETH Zurich}
}
\email{tinor@student.ethz.ch}

\author{Laurent Vanbever}
\affiliation{
	\institution{ETH Zurich}
}
\email{lvanbever@ethz.ch}


\begin{abstract}

Each year at ETH Zurich, around 100 students collectively build and operate their
very own Internet infrastructure composed of hundreds of routers and dozens of
Autonomous Systems (ASes). Their goal? Enabling Internet-wide connectivity.

We find this class-wide project to be invaluable in teaching our students how
the Internet infrastructure \emph{practically} works. Among others, our students
have a much deeper understanding of Internet operations alongside their pitfalls.
Besides students tend to love the project: clearly the fact that all of them need
to cooperate for the entire Internet to work is empowering.

In this paper, we describe the overall design of our teaching platform, how we
use it, and interesting lessons we have learnt over the years. We also make our
platform openly available~\cite{github_repo}. 

\end{abstract}


%

\maketitle

\begin{figure*}[h!]
    \centering

    \begin{adjustbox}{minipage=\linewidth,scale=0.9}

    \begin{subfigure}[h!]{0.3\textwidth}
        \centering
        \captionsetup{width=0.92\linewidth}

        \includegraphics[width=0.98\linewidth]{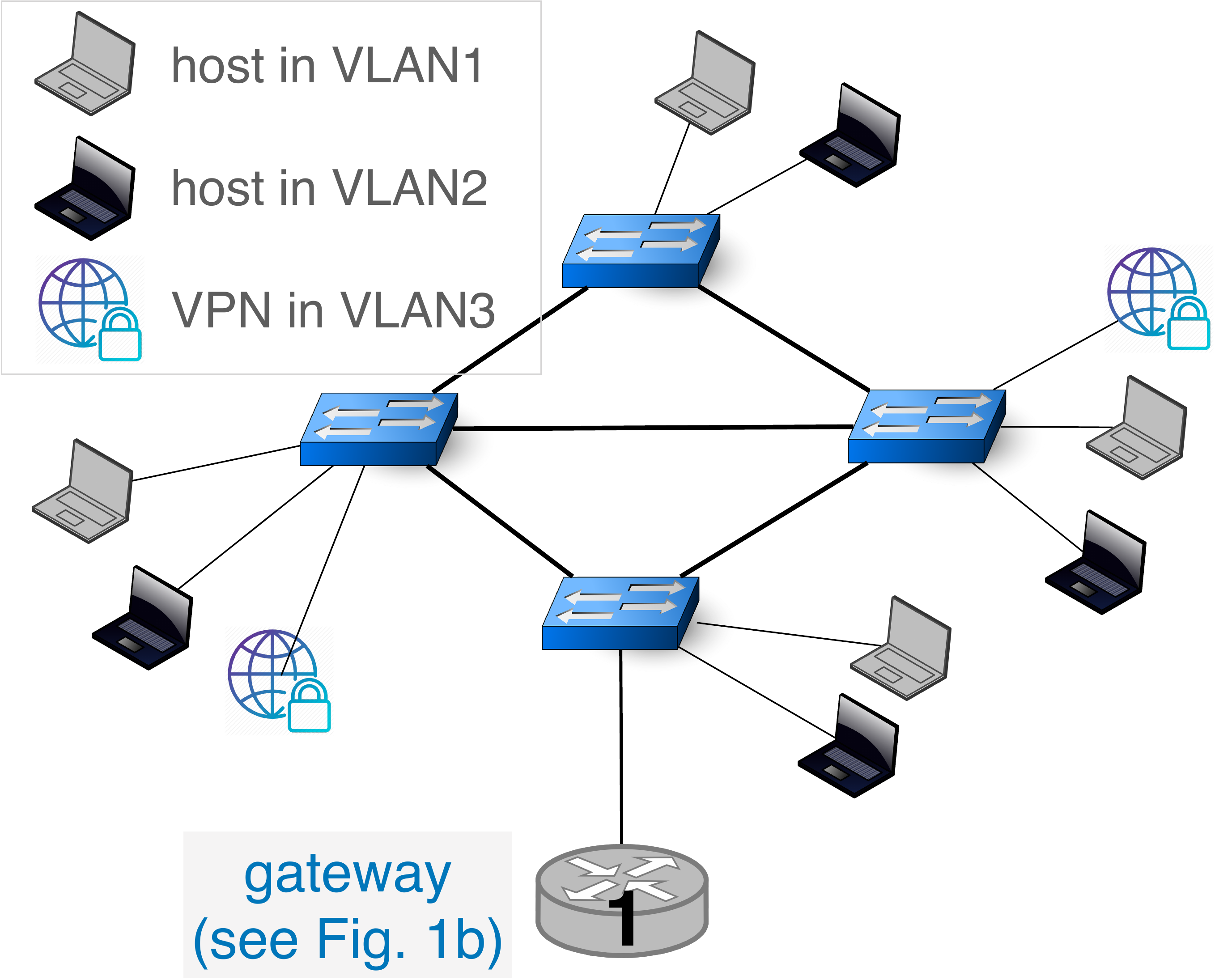}
        \caption{The L2 network. Hosts are in different VLANs, and router 1 is the gateway.
        }
        \label{fig:l2network}
    \end{subfigure}%
    \hfill
    \begin{subfigure}[h!]{0.50\textwidth}
        \centering
        \captionsetup{width=0.95\linewidth}

        \includegraphics[width=0.93\linewidth]{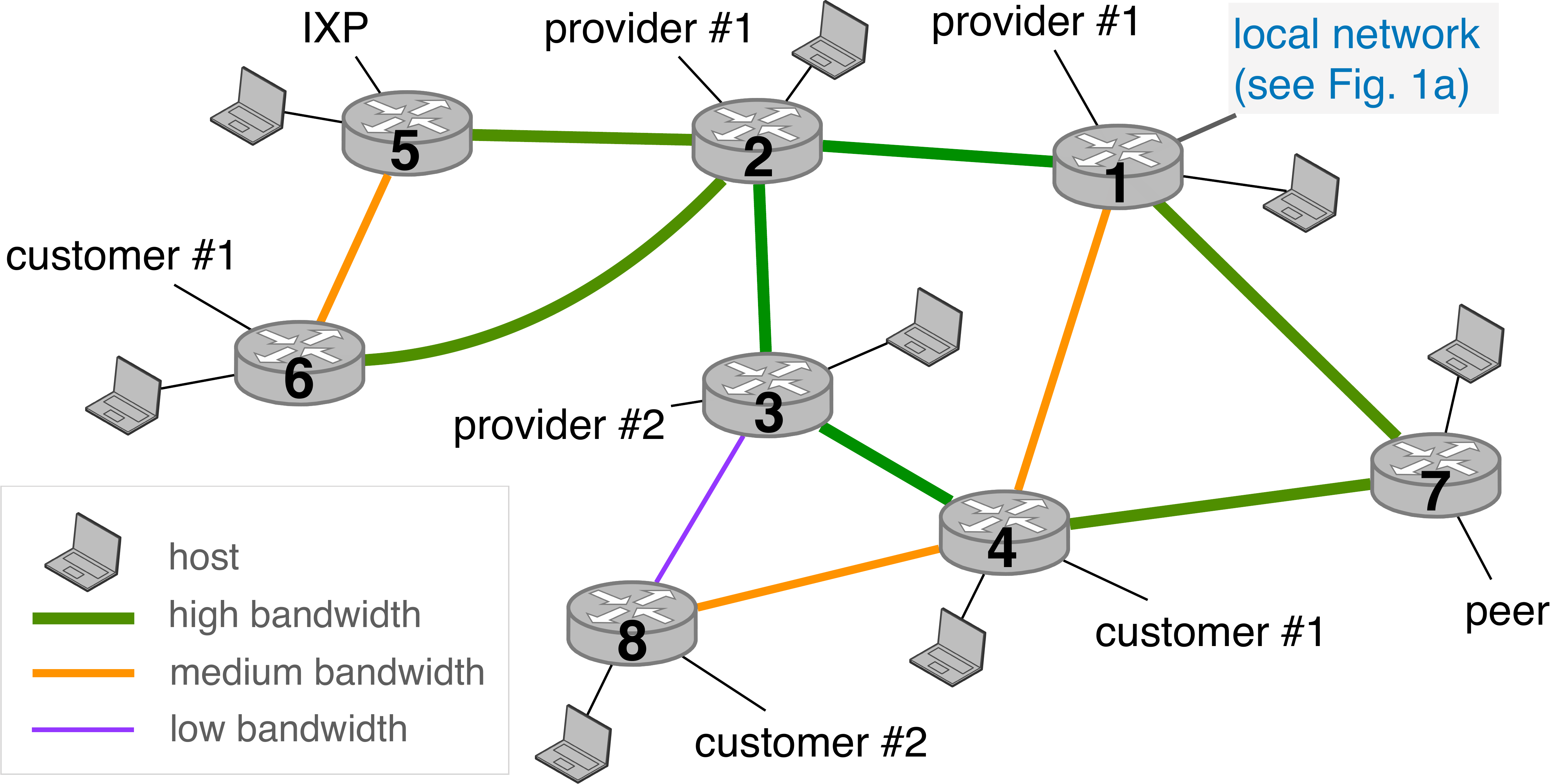}
        \caption{The L3 network. Several routers interconnect hosts and L2 networks.
        Routers also connect to other ASes or IXPs.
        }
        \label{fig:l3network}
    \end{subfigure}
    \hfill
    \begin{subfigure}[h!]{0.11\textwidth}
        \centering
        \captionsetup{width=1.1\linewidth}

        \includegraphics[width=0.95\linewidth]{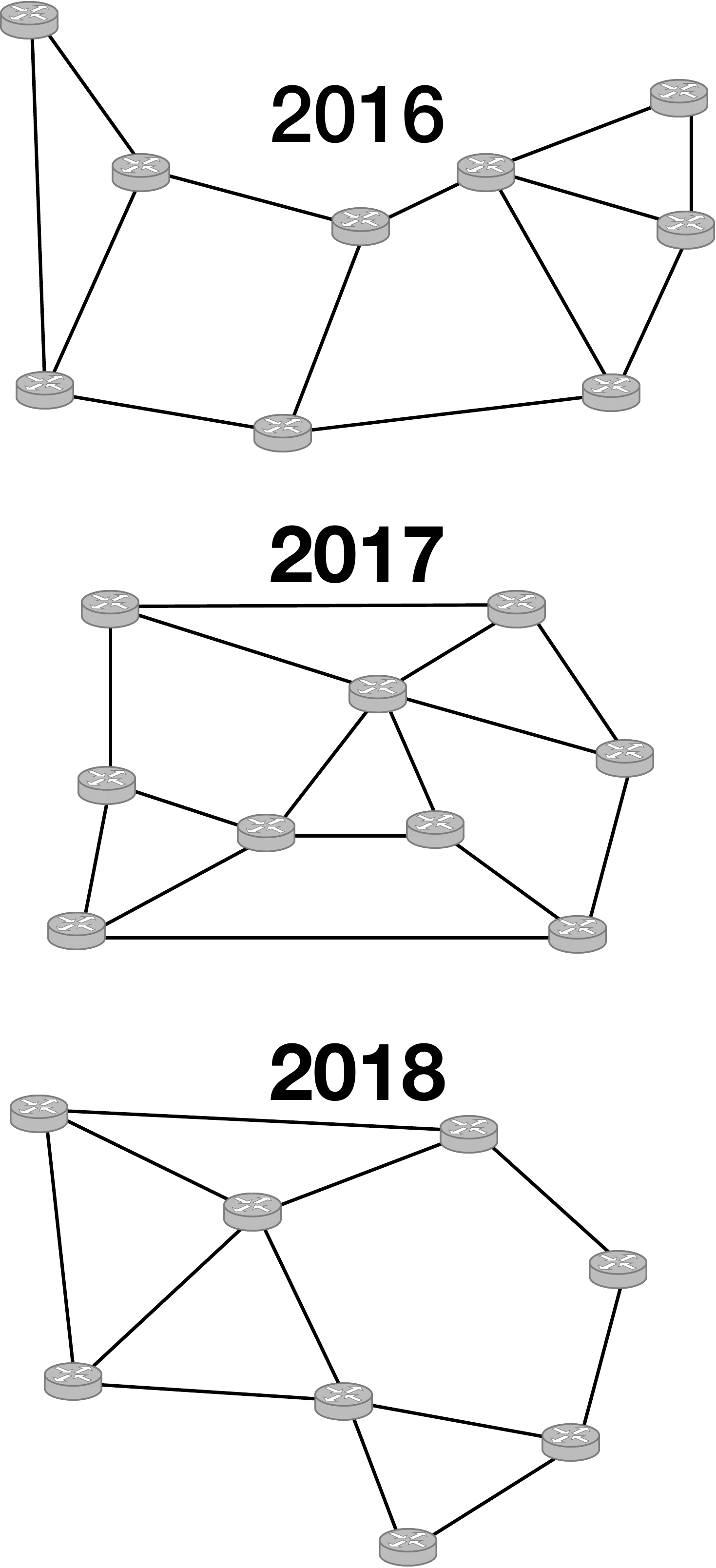}
        \caption{Previous L3 topologies.}
        \label{fig:l3network_old}
    \end{subfigure}
    \caption{In our \name project (2019 iteration), each AS has a local L2 network
    (\fref{fig:l2network}) and a L3 network (\fref{fig:l3network}).
    Every year we change the L2 and L3 topology.
    \fref{fig:l3network_old} shows the L3 topologies we used in the previous years.}
    \end{adjustbox}

\end{figure*}

\section{Introduction}
\label{ref:introduction}

Most undergraduate networking courses, including ours~\cite{eth_course},
aim at teaching ``how the Internet works''. For the instructor,
this typically means painstakingly going through the TCP/IP protocol stack, one
layer at a time, following a bottom-up~\cite{peterson2007computer} or top-down approach~\cite{kurose2005computer}. At
the end of the lecture, students (hopefully) have learnt concepts such as
switching, routing, and reliable transport; together with the corresponding
protocols.

Learning these concepts is not sufficient to understand how the Internet infrastructure works or,
alternatively, why it does \emph{not} work. For this, we think one also needs to understand the ins and outs of how the
Internet is operated which includes topics such as network
design, network configuration, network monitoring, and\dots\space network debugging.
Understanding these topics is important as Internet operations tend to have a
\emph{huge} impact. Among others, most of the Internet
downtimes are due to human-induced errors~\cite{juniper_downtime}. 

We argue that an effective way to teach students about Internet operations---one that we have successfully used for the last four years---is simply to let students operate their own \name.


\myitem{Turning students into operators.} Each year, for the last four years, around 100 ETH students have built,
configured, and monitored an actual Internet infrastructure composed of
hundreds of routers split across 60 Autonomous Systems (ASes). Each group of
2--3 students is responsible for administering, from scratch, one AS composed
of multiple hosts, layer-2 switches and layer-3 routers. Each network ``peers'' with
others using BGP, either directly or through Internet eXchange Points (IXPs),
which we (the instructors) maintain. The students' goal is identical to the
ones of actual operators: enabling Internet-wide connectivity, between any pair
of IP prefixes, by transiting IP traffic across multiple student networks. As
they quickly realize though, achieving this goal is challenging and requires a
truly collective effort. We found this to be empowering. The fact that
all networks need to work for the Internet as a whole to work really helps to
bring together the entire classroom.

Over the years, the \name project has become a flagship piece of our networking
lecture, one that the new students look forward to. Thus far, the feedback we
received from the students has been extremely positive, with comments such as:
\textit{"It really allows us to apply the theoretical concepts"}; \textit{"I am
quite confident about many things on the Internet now"}; and \textit{"It is a
unique project"}.


Besides gaining a \emph{much} deeper understanding of the various Internet
mechanisms, having students build and maintain their own Internet infrastructure
enables them to quickly realize the pitfalls and shortcomings behind Internet
operations. Students quickly realize: \emph{(i)} how fragile the Internet
infrastructure is and how dependent they are on their neighbors' connectivity;
\emph{(ii)} how hard it is to troubleshoot Internet-wide problems; and
\emph{(iii)} how difficult it is to coordinate with each other to fix remote
problems. Each year, several groups of students come up with proposals
(sometimes, even implementations!) to improve Internet operations. These
proposals often directly relate to research topics active in our community (such
as configuration verification/synthesis or active probing). Perhaps candidly, we
believe that encountering operational problems early on in their networking
curriculum can help the next-generation of network designers avoid repeating
the mistakes made in the past.





\myitem{An open platform.} Given the success of our project, we have open
sourced the entire platform~\cite{github_repo} and hope that other institutions
will start using it. We built our platform with three key goals in mind.

First, we aimed at faithfully emulating the real Internet
infrastructure. To do so, we rely on (open-source) switching and routing
software implementing the most well-known protocols (\eg STP, OSPF, BGP). We
also rely on virtualization (containers) to interconnect \emph{many} instances
(100+) of these software. While relying on virtualization in network education
is not new (\eg\cite{Baumgartner2003,Dobrilovic2008,Dinita2012,Martinez2013,Lantz:2010}),
our setting is unique as it is entirely designed to support and facilitate
large and collectively-operated routing infrastructures.

Second, while we wanted the students to learn the intricacies of Internet
operations, we also wanted to avoid making it too daunting for them. In
particular, our students only have four weeks to build the entire \name.
To help them, we developed a suite of troubleshooting tools such as a
perfect ``looking glass'' which allows them to see the routing information of
any network, together with a real-time visualization of the overall Internet
connectivity. 

Third, we wanted the setup to be easy to manage for us (the instructors),
flexible (so that we can adapt it each year), cost-effective and scalable (to
100+ students). We therefore automated the entire provisioning: it takes only a few hours
to create and launch a new \name
topology. We also optimized the setup so that it can handle 100+ students on
\emph{a single} off-the-shelf server. For larger classrooms, the project can be distributed over multiple servers.

\myitem{This paper.} 
%
After providing details on 
the capabilities of the platform (\xref{sec:project}), we show how we use it in our introductory lecture (\xref{sec:practice}),
and describe various pedagogical insights we learnt over the years (\xref{sec:lessons}).
We then highlight how we designed the platform (\xref{sec:design}), and explain how this design makes the \name suitable for various teaching objectives (\xref{sec:adapting}) and scalable to large classrooms (\xref{sec:eval}).

\section{The mini-Internet project}
\label{sec:project}

In this section, we first describe the main components (\xref{sec:architecture})
before introducing the configuration and monitoring tools (\xref{sec:tools}).



\subsection{Base components}
\label{sec:architecture}



\begin{figure}[t]
    \centering
    \includegraphics[width=0.75\linewidth]{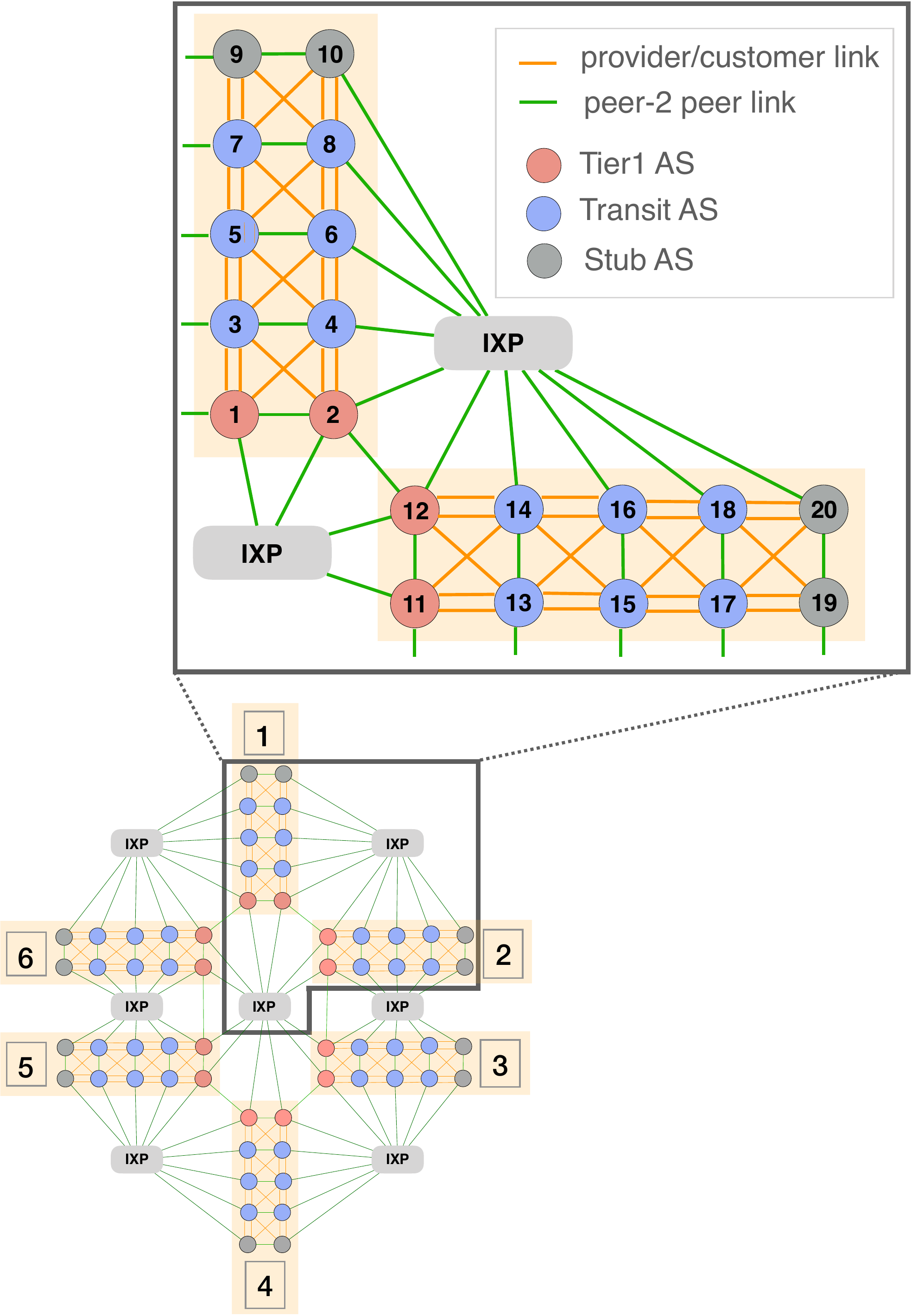}
    \caption{The topology our students operated in 2019: 60 ASes divided in six regions interconnected via seven IXPs.}
    \label{fig:aslevel_network}
\end{figure}

At the highest level, the \name is composed of several ASes
connected directly or through IXPs (\fref{fig:aslevel_network}). Each AS is
maintained by a distinct group of students and contains several routers,
switches, and hosts interconnected through links with configurable bandwidth
and delay. Each host can run tools such as \texttt{ping}, \texttt{traceroute}
or \texttt{iperf}. When the \name is correctly configured, any two hosts can
communicate with each other.

Within an AS, hosts are connected either to L2 switches or to L3 routers and can be
located in different VLANs (\fref{fig:l2network}). At least one switch is connected to a
L3 router which acts as an IP gateway. As an example, router 1 in \fref{fig:l3network}
is connected to the local L2 network depicted in \fref{fig:l2network}. L3 routers
connect to each other internally, but can also connect to routers in other ASes.

In addition, the \name supports external hosts through the use of L2-VPN
servers. Doing so enables the students to connect their own devices to their
network.

\subsection{Configuration and monitoring}
\label{sec:tools}

Similarly to the real Internet, students configure their network
devices through text-based command-line interfaces. To make this task
(slightly) less cumbersome, we also provide them with a set of monitoring tools
and services.

\myitemit{Hosts.} All our hosts run Debian Stretch~\cite{debian_stretch} and support traditional commands to measure network connectivity (\eg \texttt{ping} and \texttt{traceroute}) and configure the routing tables (\eg \texttt{ip}).

\noindent\emph{\textit{Switches and routers.}} The L2 switches are Open vSwitches~\cite{Pfaff:2015} while the L3 routers run
FRRouting~\cite{frr}. Both software suites are well-documented, support the main L2
and L3 protocols, and offer similar configuration interfaces than actual switches and routers.

\myitemit{Looking glass.} In the Internet, operators often rely on ``looking
glass'' services~\cite{looking_glass} to access the routing tables of remote ASes. Similarly, the students can access a web interface which contains periodically updated routing tables of each router in the \name.

\myitemit{Active probing.}
Network operators also use measurement platforms (\eg~\cite{ripe_atlas}) to
verify the connectivity from an external point towards their AS. In
the \name, students can run \texttt{ping} and
\texttt{traceroute} commands between any two ASes to monitor the
connectivity and forwarding paths between them. 

\myitemit{Connectivity Matrix.} Students can access a dynamic webpage which displays whether any two ASes can reach each other as a matrix. The matrix not only gives a good overview of the overall connectivity but also helps pinpointing problems (\xref{sec:orga}).

\myitemit{DNS.} Finally, we run one DNS server enabling students to use domain names instead of IP addresses.

%
%
%

\section{The \name at ETH Zurich}
\label{sec:practice}

We now explain how we use the \name in the classroom. We first explain
how we design the topology (\xref{sec:topo}), then how we organize
the project (\xref{sec:orga}) before describing what we ask the students to
do (\xref{sec:questions}).\footnote{See~\cite{github_repo} for the 2019 (and 2020) assignments.}
Finally, we explain the limitations we often encounter and how we deal with them (\xref{sec:limitations}).
%

\subsection{Topology}
\label{sec:topo}

Our implementation of the \name allows us to define the topology of the network
at every layer, i.e. L2, L3 and AS-level. \fref{fig:l2network} shows the L2
topology we used in the lecture in 2019. There are four switches, and each switch is connected to two hosts and possibly a VPN server. One switch is connected to
a gateway router. The gateway router belongs to the L3 topology displayed in
\fref{fig:l3network} which contains eight routers. In addition, one host is connected to each
router. 
\fref{fig:l3network_old} depicts the L3 topologies we used in the previous years. The topology in 2016 resembles the Internet2 topology~\cite{internet2_topo} while the one from 2018 resembles the SWITCH topology~\cite{switch_topo}.

For fairness, each student AS has the same L2 and L3 topology. \fref{fig:aslevel_network} depicts how we interconnect these ASes to
form the entire \name in 2019. 
There are 60 ASes grouped into six different regions. The topology
exhibits many of the properties found in the actual Internet: there are Tier1s, stubs and transit ASes, connected through customer/provider and peer-to-peer links. Tier1 ASes are connected in a full-mesh and several IXPs interconnect the different regions in the topology. 
Every transit AS is connected to exactly two customers, two providers, one peer and one IXP.

\subsection{Organization}
\label{sec:orga}

To reduce the student's workload, we group them in
teams of three and give each group one transit AS to operate. (We configure the Tier1s and stubs ourselves). We further allocate one /8 prefix to each AS to allocate to their hosts and interfaces.

\begin{figure}[t]
    \centering
    \includegraphics[width=1\linewidth]{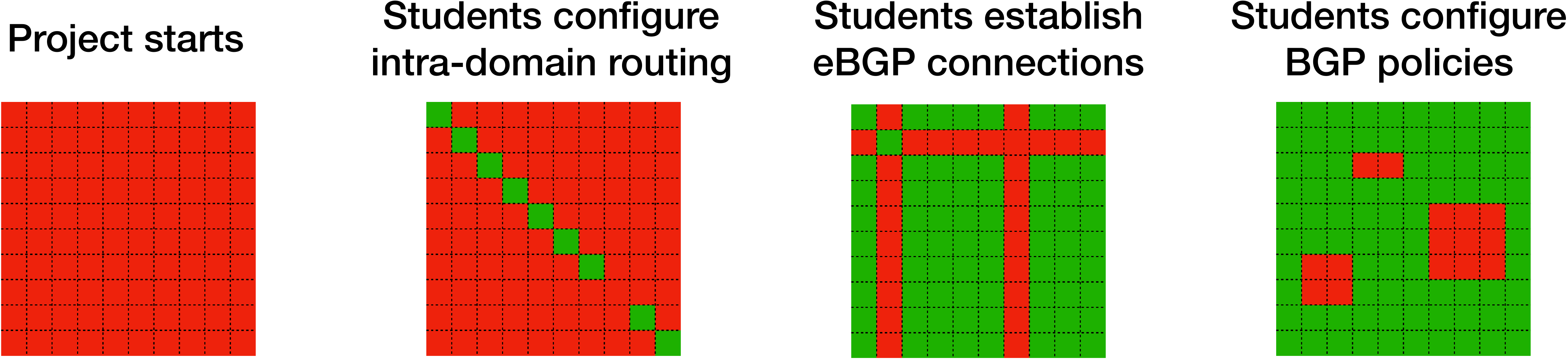}
    \caption{Snapshots of the connectivity matrix. A
    green cell indicates connectivity between two ASes.}
    \label{fig:matrix}
\end{figure}

We divide the project into the three subsequent phases: (i) establishing intra-domain connectivity; (ii) establishing inter-domain connectivity; and (iii) configuring external routing policies. These phases map to different levels of ``Internet-wide'' connectivity which we depict in \fref{fig:matrix} with connectivity matrix (\xref{sec:tools}) snapshots.

First, students have to configure the L2 switches as well as the intra-domain
routing so that hosts inside \emph{one} AS can reach each other.
As a result, the diagonal cells in the connectivity matrix should turn green.
Second, they have to configure iBGP
sessions and establish eBGP sessions with their neighboring ASes and IXPs. In the
best case, the matrix should now be completely green. Every student group can
reach every other group.
Finally, we ask the students to configure certain BGP
policies \eg to follow business relationships. At this point, the matrix
often fluctuates as students tend to make mistakes when configuring their policies.

As we can see in \fref{fig:matrix}, the matrix does not only show the current
progress but it also helps us to quickly identify mistakes. For instance, if a cell
is red in the diagonal it likely means that the corresponding group has
configuration mistakes in their intra-domain routing part. If an entire column is
red, the corresponding AS has not properly configured eBGP sessions. Finally,
asymmetric patterns often hint towards mistakes in BGP policies.

To guide students through the project, we set up a dedicated online chat room in which students can ask questions, and we organize a Q\&A session every week where several TAs provide support. We also organize a ``hackathon'' in-between the first and the second part where all students meet, discuss with their direct neighbors which IP addresses to use on their external links and figure out what is or is not working. It is also very rewarding for the students to see the matrix turning more and more green as they set up their eBGP sessions.
We also leverage the hackathon to perform a live demonstration of the effects of a BGP hijack. 


\subsection{Questions}
\label{sec:questions}

We now describe some questions we ask our students. Besides answering them, students also need to explain how they verified the correctness of their networks. Note that the students often do not have any prior networking knowledge.

\myitemit{Configure IP addresses and subnets.}
First, students must configure IP addresses and subnets for each host and router
interface. To guide them and to simplify the grading process, we ask all the groups
to follow the same scheme. For
instance, the subnet X.0.200.0/23 should be allocated to the L2 network,
where X is their AS number.

\myitemit{Configure the L2 network.}
We then ask the students to configure the local network (see
\fref{fig:l2network}) to enable direct L2 connectivity between hosts in the same
VLAN, but not between hosts in different VLANs. In the latter case the hosts
must communicate via the L3 router. Additionally, we ask the
students to configure the switch ports in such a way that the final spanning
tree follows a certain pattern.

\myitemit{Configure intra-domain routing.}
To enable connectivity within the AS, we ask them to configure OSPF network-wide. At this point, every host in one AS can communicate with any other one in the same AS.
We then ask the students to do some traffic engineering. Typically, they should
configure load-balancing across different paths. 
Additionally, we also ask them to implement routing decisions which are not achievable by simple OSPF weight changes.

\myitemit{Configure inter-domain routing.} This starts during the hackathon. Students must first configure an iBGP full-mesh before setting up the eBGP sessions with their neighboring ASes as well as the IXPs. 


\noindent\emph{\textit{Configure BGP policies.}} This is often the most complicated part. We first ask the students to configure
the local-preferences as well as the exportation rules to implement the
customer/provider and peer-to-peer business relationships with their
neighbors~\cite{Gao:2001}. Then, we ask them to implement more advanced policies
to influence the inbound or outbound traffic. During this process the students learn about the different BGP attributes (\eg AS path, MED) and how
to use them in order to influence the forwarding behavior. For instance, we often ask them to configure BGP such that the inbound traffic coming from the provider with whom they have two external BGP sessions (see~\fref{fig:l3network})
arrives preferably via one of the routers.

\myitemit{Detect and mitigate malicious operations.}
Finally, we often ask questions about malicious operations such as BGP hijacking. Typically, we would hijack a part of the IP space of each group and ask the students to identify the malicious AS before mitigating the attack (e.g., by advertising more-specific prefixes).

\subsection{Limitations}
\label{sec:limitations}

Configuring and monitoring a network is sometimes tricky, and this can become a
limitation especially given the limited time budget the students have. First,
the students are not familiar with the routers' CLI, and configuring routing
protocols is not straightforward for beginners, \eg we often have questions on
how to
configure \texttt{route-maps}. To help the students we therefore provide
additional documentation tailored to the questions we ask.


Second, some students might have persistent misconfigurations, or just start to
work on the project late. This affects the connectivity of their neighboring
ASes which for instance might not be able to reach some regions in the \name.
Debugging is therefore more difficult and some of our questions cannot be
properly answered.
To mitigate these problems, we have designed the AS-level topology (see
\fref{fig:aslevel_network}) such that each transit AS always has two providers
and two customers to prevent a network-wide loss of connectivity if one neighbor
fails. In addition, Tier1, stub ASes and IXPs are automatically configured.
This already enables the students to answer most of the questions independently
of the other transit ASes. Finally, we also adapt the grading
scheme accordingly.

\section{Lessons learnt}
\label{sec:lessons}

In this section, we describe some of the lessons we learnt over the last four years using the \name project in our lecture.

\myitem{Connectivity is a collective effort.} It is interesting to observe how
the student's perception to connectivity problems changes.
At the beginning of the project, they often show us their configuration and ask
``what did \emph{we} wrong?''. Most of the time their configuration
is correct and the problem comes from other ASes. Towards the end of the project though, they tend to first
blame other groups before searching the error source in their own
solution. As previously explained, we therefore try to build redundant topologies
such that the students do not only depend on a single
other AS. In addition, we also grade each group on their individual configurations rather than on the overall connectivity.
In conclusion, these experiences help the students to understand
that an Internet-wide connectivity requires communication between multiple
parties.

\noindent\emph{\textbf{BGP is difficult to master.}} Every year, the most confusing
topic is the interaction between the control-plane
messages (BGP advertisements) and the data plane. Most students have a hard time
to realize that they have to \eg adapt the \emph{outgoing} BGP messages in order
to influence the \emph{incoming} traffic. They often wrongly
believe route-maps are applied on each IP packet traversing the router. In
addition, some students are confused with the language used to configure the routers
as it does not follow modern programming language principles.
%
%
As a result, we improved lecture slides and
documentation and put more effort into showing the students the impact of
their configuration using \eg the measurement platform.
Overall, we hope these insights help the students in the
future to eliminate shortcoming of existing solutions should they have to
develop new protocols or ways to configure network devices.



\myitem{Automation is key.} To show the students all the required configuration steps,
we do not provide any automation tools.
Yet, certain configuration parts are shared between all
devices and could therefore be generated automatically.
%
Every year, multiple students submit simple Python or Bash scripts which automatically generate the
configuration of all devices in their network. Some students also automate the verification
process and \eg regularly ping each host in the \name. It is very encouraging to
see that the \name mimics the real Internet infrastructure closely enough such that the students can uncover actual research topics on their own (\eg configuration
synthesis and network verification).


\myitem{Visualization is important, but also dangerous.} Visualization tools such as the connectivity matrix (Fig.~\ref{fig:matrix}) are essential for the students (and network operators alike, \eg~\cite{pingmesh}) to quickly get an overview of their network connectivity. At the same time though, students often incorrectly assume their configuration is correct as soon as \eg the matrix lights up in green. As the visualization tools do not reveal \emph{all} the possible problems in the network, relying solely on them is often misleading. 
To address this problem, we plan on continuing to improve the quality of our visualization solutions in the future releases of the \name.
For example, we plan to implement a web interface that
shows the used AS path between two ASes (similarly to~\cite{tracemon}) and
highlights ASes that do not follow the business relationships.
%

\section{Implementation}
\label{sec:design}

We implemented our platform in \textasciitilde3700 lines of Bash and make
it publicly available~\cite{github_repo}. By default, our implementation runs a
\name with 20 ASes. We provide various configuration files, e.g. to reproduce
the L2 and L3 topology depicted in \fref{fig:l2network}
and \fref{fig:l3network}.
However the topologies can easily be customized.

In this section, we give more details on how we build the virtual networks, implement
the various monitoring and debugging tools and explain how the students can access
the \name.

\myitem{Building the network.}
We build the \name with Docker containers~\cite{Merkel:2014}. As opposed to
virtual machines, a container does not run its own operating system, but relies
on namespaces, a feature available in the Linux kernel. Namespaces isolate
software from its environment by partitioning kernel resources. Docker
containers are lightweight because they share the host machine's system kernel
and computational resources are dynamically allocated.
Each component in the \name (hosts, switches and routers) runs in its own
dedicated Docker container. We then connect the Docker containers following the
\name topology using Open vSwitch~\cite{Pfaff:2015} bridges and virtual ethernet
links.
The containers run Debian Stretch~\cite{debian_stretch} and we add the main
networking tools (\eg \texttt{traceroute}, \texttt{dig}).
For the switches, we also use Open vSwitch, a software switch which
supports VLANs and the Spanning Tree Protocol. For the routers, we deploy
FRRouting~\cite{frr}, an IP routing suite which uses the native Linux/Unix IP
networking stack and supports the main routing protocols.

We use
OpenVPN~\cite{openvpn} to allow the students to virtually connect an external client (e.g., their laptops) into the \name. The OpenVPN processes run in the server hosting the
\name and are connected to the \name with virtual links. Each of them listens for new connections
on a different port belonging to the host server interface which is connected to the actual Internet.
By choosing a specific port, students can therefore decide where to connect in the topology.


\myitem{Setting up monitoring and debugging tools.}
For the looking glass, we automatically pull the routing table of each router
every minute and upload them to a website. For the measurement platform, we use
a dedicated container, connected to every AS and accessible by all the students,
from where they can run measurements. Two additional containers are created, one
is dedicated for the connectivity matrix and the other one for the DNS service.
These two containers are connected to every AS but are not accessible by the
students. The container used for the connectivity matrix performs ping
measurements at regular intervals between all the pairs of ASes and the results
are uploaded to a webpage. For the DNS service, we automatically generate the
configuration file and run a \texttt{bind9}~\cite{bind9} server in a dedicated
container.

%
\myitem{Isolated student access.}
Our students should be able to easily access all their network devices but must
not have access to containers of any other group.
To achieve that, we rely on the natively provided isolation of Docker containers
as well as SSH connections.
More precisely, we deploy one additional container for each group of students
that we use as a ``proxy'' and tunnel the incoming SSH connections to the
corresponding proxy container based on the port number. We allocate one port
number to each group and share the SSH password for a given proxy container
only with the students of the corresponding group. From a proxy container, a
student can then easily jump into the CLI of one of his or her virtual devices
using a simple script that relies on SSH and public/private key pairs
automatically generated during the \name startup process.
The following commands illustrate how to access router 3 in AS1 (port 2001 is
allocated to the proxy of AS1):

\shellcmd{ssh -p 2001 root@server.ethz.ch} \\
\shellcmdproxy{./goto.sh 3 router \hfill    \textcolor{gray}{\textit{\# Could also be "host"}}} \\
\routercmd{show ip bgp}

\noindent
Observe that the students can setup a key-based SSH authentication to simplify
the access to the proxy container.

%

%

\section{Adapting the mini-Internet}
\label{sec:adapting}

This section highlights several ideas on how instructors can increase the authenticity of the \name infrastructure as well as the difficulty of the questions (e.g., for more advanced classes).



\myitem{Adapting the topology and the monitoring tools.} Instructors can easily
tailor the L2, L3 and AS-level topology of the \name to their requirements.
Among others, they can adapt the number of routers within each AS or the
number of ASes. A larger number of routers raises interesting scalability
questions which opens new possibilities to teach about e.g., hierarchical
routing.


To make the \name even more realistic, instructors can introduce latency on
certain links to mimic \eg the geographical location of certain ASes and IXPs. In addition, instructors could also fail links and/or routers while the project is running allowing students to test their network resiliency. Instructors could furthermore deploy new monitoring tools (such as BMP~\cite{rfc7854}) or modify the existing ones to only show partial information (\eg by configuring ICMP filtering). More generally, whatever feature the underlying software tools (\eg FRRouting or Open vSwitch) support can also be used in the \name.

\myitem{Adapting the questions.}
Besides configuring additional protocols such as IPv6 or MPLS, instructors could also completely change the structure of the questions and the overall
teaching goals.
To list a few examples, one could confront students with a ``blackbox'' network where they first have to use measurement tools (\eg \texttt{traceroute}) to figure out and visualize the topology of their network as well as the interconnections with other student groups.
Another idea relates to the grading methodology. Instead of grading students based on the correctness of their configurations, one can introduce a virtual currency and ``bill'' the students according to the amount of traffic transiting through their network depending on the business relationships with their neighbors. Yet another idea would be to rely on the \name to train students or network operators to correctly use and implement emerging technologies such as to validate the origin of BGP routes using the RPKI infrastructure~\cite{rfc6480}.

\section{Evaluation}
\label{sec:eval}

We now show that our platform is well-suited to be used as a practical project
in computer network courses with 100+ students.\footnote{In 2020, we successfully used the mini-Internet for 150 students.} We evaluate it on
an Ubuntu 18.04.3 server with 24 Intel Xeon CPU
cores @ 2.30\,GHz, 256\,GB of memory and running the 4.15.0 Linux kernel. We
always fully configure hosts, switches and routers and use
the 2019 topologies depicted in \fref{fig:l2network} and \fref{fig:l3network}.
For tests with 60 ASes, we use the topology in \fref{fig:aslevel_network}.
For topologies with 20 or 40 ASes we keep the same AS-level structure
but reduce the number of regions accordingly \eg we use two regions to
form a \name with 20 ASes.

\myitem{The \name is easy and relatively fast to setup.}
\begin{figure}[t]
    \centering
    \includegraphics[width=0.7\linewidth]{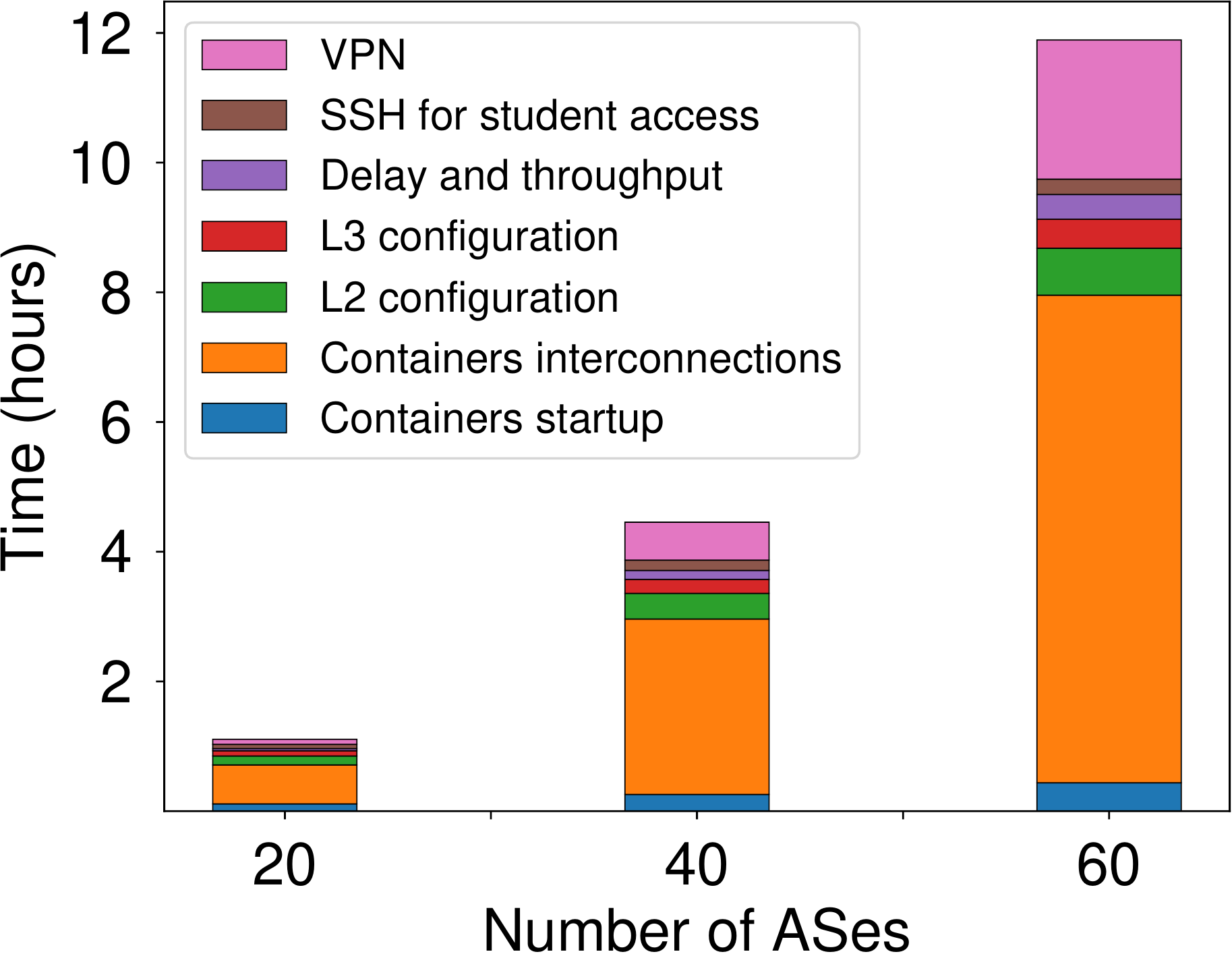}
    \caption{Startup time vs the number of ASes.}
    \label{fig:startup}
\end{figure}%
To start the \name, the instructor only has to define the topology in the
configuration files and run a Bash script. \fref{fig:startup} reports the
startup times depending on the number of ASes in the \name. We can see that for 60 ASes \ie the size we used in 2019 at
ETH Zurich, it takes around 12 hours to build the \name. This is acceptable given that this step is done automatically and only once at the beginning of the semester. Digging deeper, we see
that the interconnection of the containers with Open vSwitches and virtual links
has the longest setup time. With 60 ASes it takes 7.5 hours to
create the 497 Open vSwitches and the 7191 virtual links used to connect the
1690 containers.
Similarly, enabling the VPN service also takes time as it needs a lot of virtual
links and we have to generate a set of keys and certificates, one for each VPN
server.

\begin{figure}[t]
 \centering
 \begin{subfigure}[t]{.23\textwidth}
	 \includegraphics[width=\textwidth]{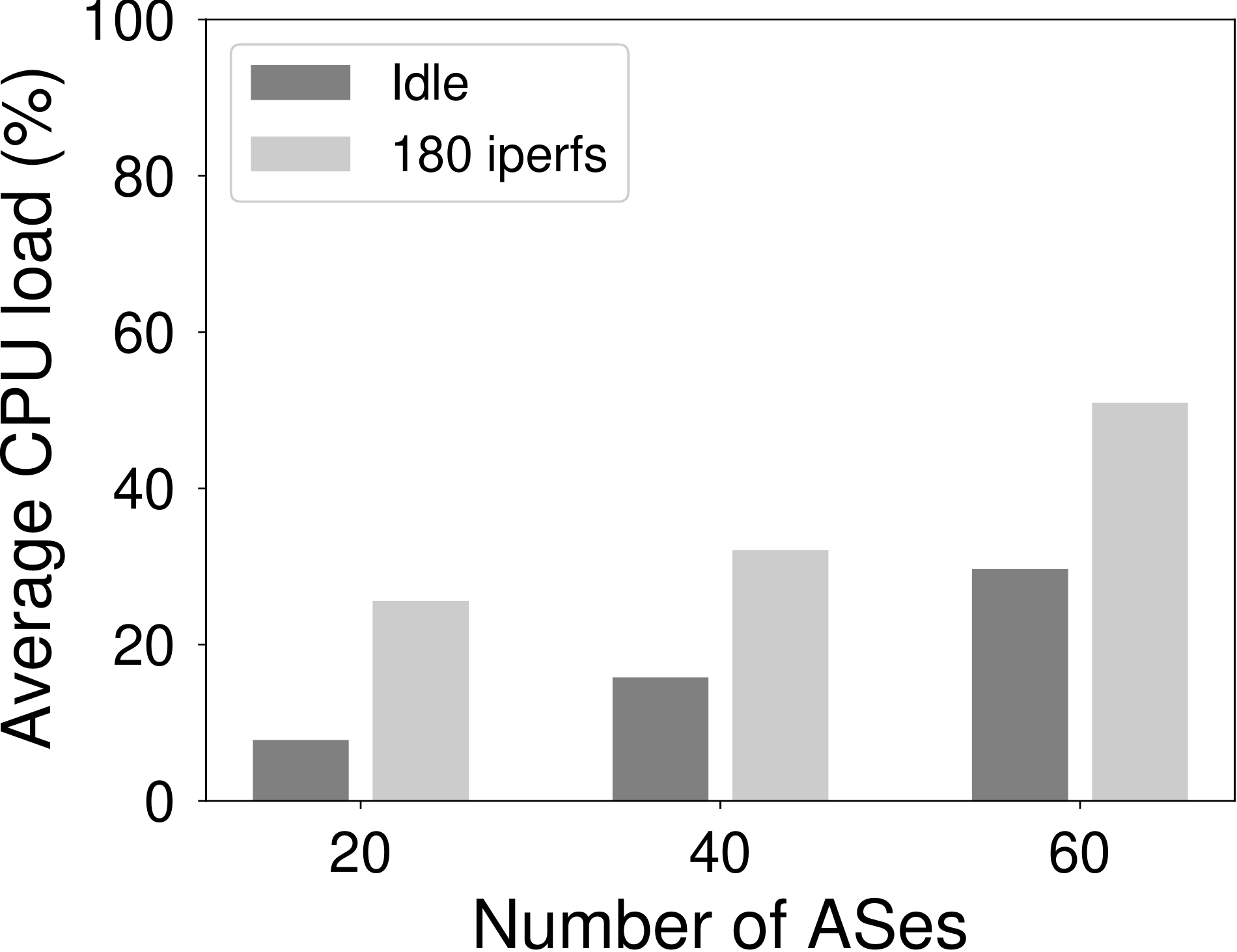}
	 \vspace*{-3mm}
	 \caption{CPU load}
	 \label{fig:cpu_load}
 \end{subfigure}
 \hfill
  \begin{subfigure}[t]{.23\textwidth}
  	 \includegraphics[width=\textwidth]{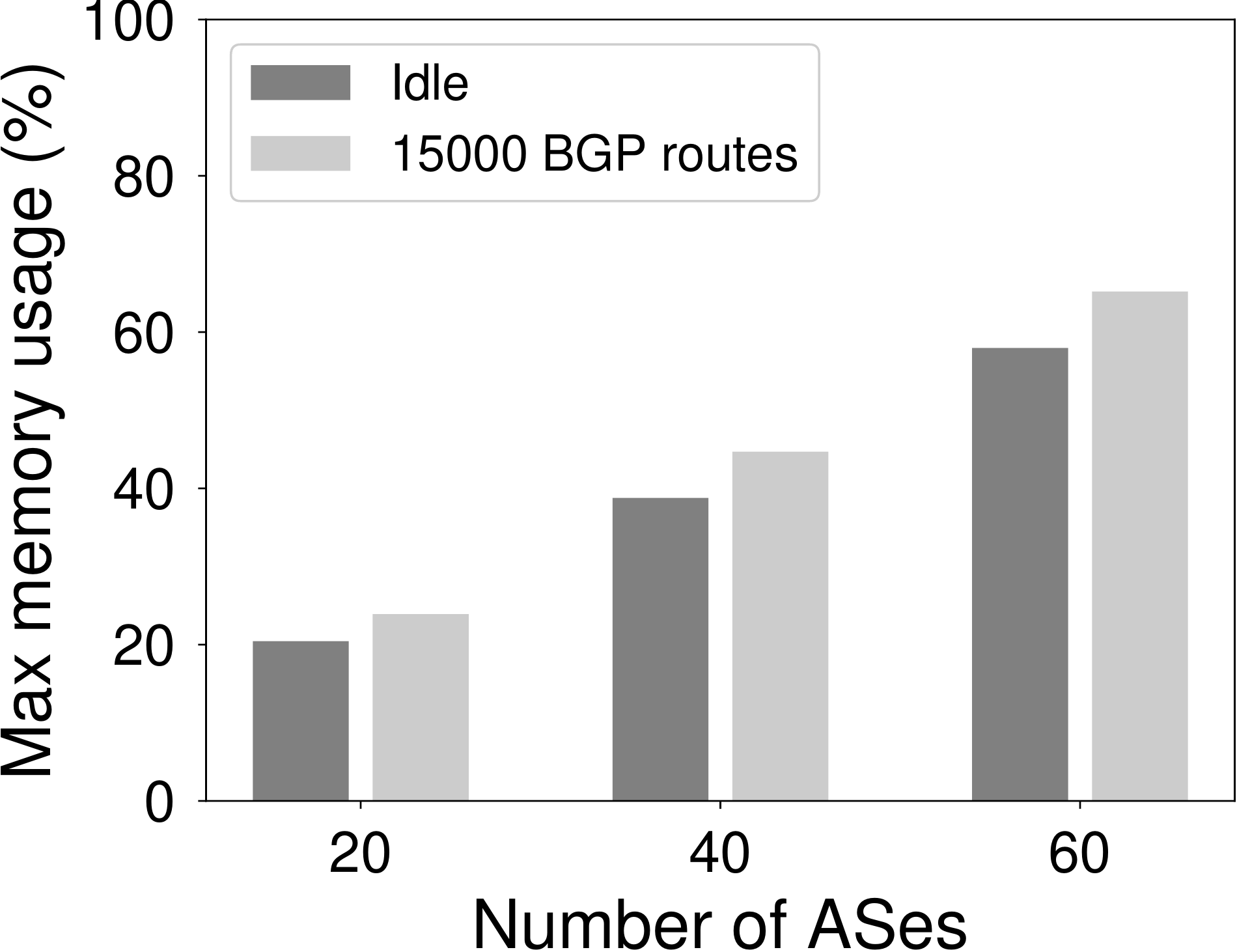}
 	 \vspace*{-3mm}
 	 \caption{Memory usage}
  \label{fig:mem_load}
  \end{subfigure}
 \caption{The CPU (\fref{fig:cpu_load}) and memory (\fref{fig:mem_load}) used by
        a \name with 60 ASes when idle and under load.}
 \label{fig:load}

\end{figure}

\myitem{One server is enough for 100 students.}
\fref{fig:load} depicts the CPU and memory usage as a function of the size of the \name.
In the idle state \ie the \name is fully started but no traffic is being forwarded, the average CPU load is 29.7\% and up to 58\% of the memory is used (topology with 60 ASes).
To simulate the network under load, we perform
two tests. First, we start 180 ten-minutes \texttt{iperf} sessions between random pairs of
hosts. In the case of 60 ASes, each student group therefore
simultaneously sends traffic to three other groups on average. Second, we also measure the load when we advertise a high number
of BGP prefixes in the entire \name. This test is based on the observation that
students often advertise more prefixes than expected. For example, some groups
advertise every single used /24 prefix instead of only their /8 prefix.
Therefore we measure the load after advertising 15000 prefixes (250 distinct BGP prefixes per group with 60 ASes).


Because all the virtual links are bandwidth limited, the \texttt{iperf} sessions do not
overload the server. The CPU load only increases to 51\% with 60 ASes (see \fref{fig:cpu_load})
whereas the effect on the memory is negligible.
The 15000 BGP routes lead to a high CPU load (>80\%) during
the convergence time, and an additional memory usage over time (up to 65.2\% with 60
ASes, see \fref{fig:mem_load}). The
results thus indicate that one server can easily handle a \name with 60 ASes,
enough for 108 students if we allocate three students to each transit AS (see
\xref{sec:practice}). Although we never had issues during the last four years,
we note that a malicious student group could probably overload the server and
impact part of the \name (\eg by advertising hundreds of thousands of fake BGP
routes). Yet, we mitigate the potential impact by periodically and automatically
saving the configuration files of each router and switch in the network.
Therefore the student's progress is not lost should we have to restart one or
multiple containers. In addition, we could also maintain logs to detect
malicious activities.

%
%
%
%
%
%
%
%
%
%
%
%
%
%
%
%

\section{Conclusion and Future work}
\label{ref:future_work}

We propose to teach students how the Internet \emph{practically}
works by having an entire classroom build and operate their own Internet
infrastructure. We describe the design and the implementation of a platform
that can support this and make it publicly available. Our four year-long
experience running the project at ETH Zurich tells us that not only do the
students like the project, they also gain a much deeper understanding of the various Internet mechanisms.

We have been nurturing the project over four years and intend to continue to do so. Below is a list of improvements we plan to do.

\noindent\emph{\textbf{Support for multi-server deployments.}\enskip} In some classes the number of students can easily exceed 100, in which case a single server might not have enough resources to run the entire \name. We intend to soon support the ability to run the \name over multiple servers, in a transparent manner. 



\myitem{Auto-grading.}
Manually grading the students by carefully checking the configuration
files of their routers and switches is time-consuming.
Here, we intend to develop tools to parse their configuration files and actively send traffic through the network to automatically check the correctness of the configurations and verify the implemented policies at runtime.
%

\myitem{Connecting the real Internet to the \name.}
We plan to connect the real Internet to the \name to enable students to browse
the web or watch videos from a host inside the \name. This must be done
carefully as not all the Internet prefixes can be advertised in the \name (some
prefixes are already allocated in the \name itself), and the additional load
must be tightly controlled (traffic volume and number of prefixes).

\remove{
\myitem{The \name as a service.}
Not every university can dedicate a server for a course project and is willing
to operate a \name instance. We thus plan to offer the \name project as a
service, where the \name runs in one of our servers at ETH Z{\"u}rich and
students from other universities can use it remotely. Students from different
universities could even work together on the same \name to enable network-wide
connectivity.
}

\section*{Acknowledgements}

We thank all the teaching assistants from the Networked Systems Group at ETH
Zurich that were involved in this project since 2016. We are also grateful to
the SIGCOMM CCR anonymous reviewers for their insightful comments. We would
also like to thank Ethan Katz-Bassett and Oliver Hohlfeld for their insightful
feedback, and for having used \emph{and} improved the platform themselves.

This work was partially supported by a Swiss National Science Foundation Grant (Data-Driven Internet Routing, \#200021-175525).

\section*{References}
{

    	\bibliographystyle{ACM-Reference-Format}
    	\bibliography{biblio}
    %
    %
}

\end{document}